\documentclass[letterpaper,times]{IONconf}

\usepackage{amsmath,amssymb}
\usepackage{graphicx}
\usepackage{url}
\usepackage{natbib}
\renewcommand{\cite}{\citep}
\usepackage[hidelinks]{hyperref}
\usepackage{enumitem}   

\title{Improved Frequency Tracking with Adaptive Moments for Narrowband Interference Mitigation in GNSS}

\author{
    Burak~Soner$^{\dagger,\ddagger}$, Abdulkadir~Uzun$^{\ddagger}$, Ekin~Uzun$^{\ddagger}$%
    \vspace{1mm} \\%
    $^{\dagger}$\textit{sobu Labs, Ankara, Türkiye} \\
    $^{\ddagger}$\textit{EDGE Microwave, Istanbul, Türkiye}
}

\begin{document}

\maketitle

% biography section. The * indicates a section excluded from numbering.
\section*{biography}

\biography{Burak Soner}{received his B.Sc.~in mechatronics from Sabancı University and his Ph.D.~in electrical and electronics engineering from Koç University, both in Istanbul, Türkiye. He is a multidisciplinary engineer with more than ten years of simultaneous research and professional experience in power electronics, control, wireless communication and sensing, optics, embedded systems, FPGAs, and artificial intelligence (AI) for edge devices. He recently founded his own company, \emph{sobu Labs}, in Ankara, Türkiye, providing research \& engineering services for compact DSP / AI algorithms on FPGAs and other SoCs, primarily for autonomous navigation applications, for customers in Türkiye, EU and USA.}

\biography{Abdulkadir Uzun}{received his B.S. and M.S. degrees from Sabancı University, Istanbul, Türkiye, in 2017 and 2020, respectively. He is currently pursuing a Ph.D. degree at Sabancı University, focusing on orbital angular momentum (OAM) antennas. He has worked as an electronic warfare system design engineer at ASELSAN, an RF and microwave engineer at ERA RF Technologies on Ka-band SATCOM antennas and transponder systems, and later at EDGE Microwave, Istanbul, on GNSS and CRPA antenna development. His research interests include OAM and GNSS antennas, Ka-band SATCOM systems, indoor positioning using GPS signals, and RF/microwave circuit design. He is a Member of the IEEE Antennas and Propagation Society.}

\biography{Ekin Uzun}{received his B.Sc.~in electrical and electronics engineering from Sabancı University and his M.Sc.~in defense technologies from Istanbul Technical University (İTÜ). He has more than fifteen years of industry experience spanning RF hardware design, microwave systems, and mission-critical electronics for aerospace and defense platforms. He began his career at Pavotek (2010–2015), where he led hardware development across commercial and military projects, and continued at CTech (2015–2022) as a lead RF and hardware designer contributing to advanced airborne and space systems. From 2022 to 2024 he served as the RF and microwave design manager at \emph{ERA RF}. He is currently a co-founder of \emph{EDGE Microwave}, Istanbul, Türkiye, developing resilient GNSS technologies and compact anti-jamming CRPA solutions.}

\section*{Abstract}

Personal privacy devices (PPDs) typically emit strong tones or swept narrowband signals to jam nearby GNSS receivers and deliberately cause loss of lock. Excision methods deployed on receivers mitigate such interferers by tracking their instantaneous frequency and removing those components in either the time domain (e.g., notch filtering) or a transform domain (e.g., Fourier-domain excision). For effective mitigation without degrading the GNSS signal, the excision location must be precise; misplacement can sometimes harm performance even more than leaving the interferer unmitigated. Trackers are therefore evaluated for both dynamic tracking performance, especially against fast-sweeping jammers, as well as steady-state estimation jitter, reflecting a fundamental tracking speed versus variance trade-off. We propose a new frequency tracking algorithm that provides a better trade-off than existing methods using first and second moments of the gradient for a complex first-order IIR notch filter under a standard output power minimization objective. We first describe the algorithm, which is inspired by the Adam optimization method widely used for optimizing neural network parameters, and its relation to existing adaptive notch filter (ANF) update rules. Next, we analyze performance over a wide range of simulated and recorded interference events, including publicly available datasets, and benchmark against state-of-the-art methods including frequency-locked-loop (FLL) based designs. Finally, we analyze resource utilization and latency, and quantify the effects of quantization and pipeline delays on the proposed method and on representative state-of-the-art baselines. Our proposed Adam-based method shows marginally higher resource usage than a vanilla ANF while providing better suppression under most scenarios.

\section{Introduction}

Global Navigation Satellite System (GNSS) receivers are vulnerable to radio-frequency interferences (RFI), which are either deliberately transmitted by jammers (e.g., military) or created via unintended transmissions from other RF equipment (e.g., communication, RADAR, or other sensors). One class of RFI that is particularly harmful and is unfortunately becoming increasingly common is strong narrowband emissions that appear as tones or sweeping chirp-like signals, typically emanating from so-called personal privacy devices (PPDs) \cite{eu_sota_rfi_2023, vandermerwe_ppd_eval_2018}. When such an interfering signal falls within GNSS processing bands, it can cause rapid loss of lock, raise tracking jitter, and degrade position, velocity, and timing (PVT) performance \cite{borio_benchmark_2020}. 

A well-studied mitigation strategy against RFI from PPDs is \emph{excision}, i.e., detecting and removing or strongly attenuating the interfering component, either in the time domain (e.g., notch filtering \cite{nehorai_anf_1985}) or in a transform domain (e.g., attenuating bins in the short-time Fourier transform (STFT) \cite{vandermerwe_hddm_2021} or other time–frequency representations like the Wigner-Ville distribution \cite{borio_tfr_excision_2008}). The effectiveness of the excision operation depends on two factors: first, the ability to accurately locate and characterize the RFI; second, the capability of the excision mechanism to effectively remove or attenuate it. For instance, for narrowband interference signals, the simplest first-order complex IIR notch filter is a natural choice for time-domain excision \cite{borio_2006_1polefilter} since it is the approximate inverse of the interferer signal, completely removing it when placed at the correct frequency. The main challenge for this time-domain excision method is then precise placement and tracking of the typically non-stationary interferer, because any misalignment of the notch decreases useful GNSS signal energy and/or leaves residual interference signals that propagate into the subsequent correlator and tracking stages \cite{gamba_postcorr_impact_2020}.

Tracking strategies that estimate instantaneous RFI frequency for proper notch placement can be categorized into two groups\footnote{Other methods exist, such as pilot notches to steer the main notch \cite{piloted_anf}, or learning-based estimators (e.g., RBF, DNN) that excel at very low SNRs, but these are too complex for typical resource-constrained GNSS anti-jamming platforms \cite{elmelhi_2014_rbf_anf, izacard_2021_deepfreq, leung_2023_learnedfreqestFRI}.}: 
\begin{enumerate}[label=(\roman*)]
    \item \emph{gradient-based} approaches, including adaptive notch filters (ANFs), which iteratively solve a non-convex problem of minimizing output power to detect and track narrowband interference in real time, and
    \item \emph{control-theoretic} approaches, such as frequency-locked-loop (FLL) designs, which use a feedback loop to track the interferer’s instantaneous frequency.
\end{enumerate}
Gradient-based methods, specifically ANFs, have garnered significant attention because they are very effective even with the simplest least mean squares (LMS) gradient-descent based adaptation mechanism, enabling digital implementation with very low latency and modest resource demand \cite{vandermerwe_ppd_eval_2018}. Their performance is governed by a fundamental trade-off: faster adaptation allows the notch to follow rapidly moving interferers such as fast-sweeping chirps, but aggressive updates worsen steady-state estimation variance and increase the risk of notch misplacement \cite{borio_loop_analysis_2016, gamba_2019_anfparametrization}. As a result, a wide variety of update laws and objective functions, as well as control-loop based methods, have been explored.

ANFs have early foundations in adaptive IIR notch-type structures \cite{nehorai_anf_1985, strobach_1995_singlesectionANF, ferdjallah_1994_anf, hush_1986_adaptiveIIR}. Projections of these designs in the GNSS space include the simplest complex first-order IIR \cite{borio_2006_1polefilter}, higher-order/multi-pole extensions \cite{borio_2008_multipole_anf} and alternative realizations such as all-pass based notch filters that retain advantageous phase properties \cite{ma_2005_allpassANF}. Objective functions of these adaptive filters generally aim to minimize some form of output power assuming that the interferer has greater in-band power compared to the GNSS signal. Various approaches include the output of the moving-average (MA) part of the IIR notch filter \cite{borio_loop_analysis_2016}, the filter output as a whole, i.e., MA + autoregressive (AR) output \cite{borio_2006_1polefilter}, nonlinear formulations such as MA$\times$AR \cite{punchalard_2008_nonplain_errorcriteria}, $p$-power criteria to reshape the error landscape \cite{pei_p_power_1993}, as well as modified/unbiased gradient formulations to mitigate bias in higher-order IIR filters \cite{loetwassana_2010_unbiased_gradient}. For a given filter structure and objective function, common parameter update rules include first-order LMS variants such as the plain one \cite{borio_loop_analysis_2016}, variable-step, or power-normalized (NLMS) variants \cite{calmettes_2001_nlms_optimumrx, cascosanchez_2011_vss_nlms}, and second-order methods such as least squares solutions \cite{zhu_2016_anf_rls_lattice, he_2018_frls_anf_fpga} or Kalman-filter based trackers \cite{kim_2019_kalman_cphd}. Recent state-of-the-art solutions against fast sweeps/chirps and PPDs also employ FLL-based designs \cite{gamba_2018_fll4chirp_anf_analysis}, and hybrids like AFLL–ANF \cite{vandermerwe_afllanf_2021} or multi-parameter formulations \cite{vandermerwe_mpanf_2022}, which are algebraically equivalent to ANFs for the same problem \cite{borio_loop_analysis_2016}. 

Our proposed technique is an improvement on the ANF, using the standard output power (MA+AR) objective but with adaptive-moment based updates, and we compare it with state-of-the-art examples of both gradient-based and FLL-based methods. Use of adaptive moments in optimization was popularized by the Adam optimizer in machine learning, which combines exponential filtering based first- and second-moment estimates of the gradient to accelerate convergence without increasing steady-state variance. While such ideas have been applied in adjacent signal processing contexts (e.g., frequency-domain equalization for mode-division multiplexing \cite{fiber_adam_anf}), they have not, to our knowledge, been studied for narrowband GNSS interference frequency tracking and excision in ANFs, especially for the chirp-type signals employed in PPDs.

\textbf{Our Contributions:} Building on these observations, we propose an Adam-inspired frequency tracker for a complex first-order IIR notch filter with a plain output power minimization objective function. Specifically, we:
\begin{itemize}
    \item derive an ANF update rule that leverages exponentially filtered first and second moments of the stochastic gradient (Adam-ANF), and describe its relationship to classical ANF update rules,
    \item simulate dynamic and steady-state performance of our Adam-ANF and compare that with the state-of-the-art under recorded and simulated RFI from our test harness as well as recently published datasets such as \cite{mehr_2025_atlantis_dataset}, and
    \item quantify resource use, pipelining latency, and quantization impacts for Adam-ANF, and compare that to the classical ANF.
\end{itemize}
Collectively, our results indicate that adaptive-moment updates can deliver stronger dynamic tracking and reduced jitter for a wide range of narrowband interferers, with marginal additional resource cost compared to a classical ANF. It is important to note that Adam-ANF does not dictate the excision method to be a notch filter, or obligate the use of a certain objective function. It is merely a better optimization method, i.e., a better tracker of instantaneous frequencies for narrowband interferers. Therefore, it is possible to combine it with any excision method or with alternative objective functions, possibly chosen with respect to a criteria of minimal distortion to the useful GNSS signal during operation.

The rest of the paper is structured as follows: Section II provides the signal model for narrowband interferers, presents the time-domain excision methods considered in this paper, and formulates the instantaneous frequency tracking problem in narrowband RFI mitigation. Section III presents the proposed Adam-based instantaneous frequency tracking method. Section IV demonstrates the performance of the proposed method and compares that with a benchmark of state-of-the-art alternative methods used in ANFs as well as analyzing resource utilization and performance hits under finite-precision quantization and pipelining delays. Section V concludes the paper with a discussion of the results and future work.

\section{Narrowband Interference Excision for GNSS Signals}

This section presents the received signal model for narrowband interference, describes the time–domain excision mechanism we consider for this study, i.e., a first-order complex IIR notch filter, and frames the instantaneous frequency tracking problem as well as the related error metrics. Our analyses focus on continuous-wave (CW) tones and chirp interferers, as these are the dominant emissions observed in personal privacy devices (PPDs) \cite{vandermerwe_ppd_eval_2018}.

\subsection{Signal Model}

The discrete-time received complex baseband signal $r[n]$ is expressed as
\begin{equation}
    r[n] \;=\; s_{\text{GNSS}}[n] \;+\; i[n] \;+\; \eta[n],
    \label{eq:signal_model}
\end{equation}
where $n$ is discrete time, $s_{\text{GNSS}}[n]$ is the GNSS signal, $\eta[n]$ is additive noise, and $i[n]$ is a narrowband interferer modeled as
\begin{equation}
    i[n] \;=\; A[n]\,e^{j\phi[n]}, 
    \qquad 
    \phi[n]=\psi + 2\pi f_i[n] n,
\end{equation}
where $f_i[n] \in (-F_s/2,F_s/2]$ is the instantaneous digital frequency, $F_s$ is the sampling rate, $\psi$ is an arbitrary phase offset included for completeness, and $A[n]$ is the amplitude of the interfering signal. $A[n]$ is constant or slowly varying with respect to signal time in the GNSS interferer context, where variation is due to channel dynamics or frequency-dependent receiver front-end artifacts (e.g., a bandpass filter enveloping a cross-band chirp). We consider two classes of narrowband interferers:
\begin{equation}
    \text{CW tone:}\quad f_i[n] = f_0, 
    \qquad 
    \text{Chirp:}\quad f_i[n] = f_0 + \nu(\dot{f_i},f_0,n),
\end{equation}
where the chirp class has subcategories based on the type of the sweep as defined by the function $\nu$, e.g., $\nu(\dot{f_i},f_0,n) =  n\,\dot{f_i}$ for linear and $\nu(\dot{f_i},f_0,n) = f_0 (\exp(n\,\dot{f_i}/f_0)-1)$ for some type of exponential sweep, with modulo operators on $n$ to facilitate repeating sweeps. In this formulation, sweep speed is determined by the magnitude of $\dot{f_i}$ relative to the sampling rate of the receiver and the update rate of the frequency tracking mechanism used by the interference mitigation method.

Although we define it for chirps, more exotic $\nu$ functions such as chirp-tick or random hopping are possible, so the space of possible $\nu$ functions effectively spans the general class of frequency-modulated continuous wave (FMCW) signals \cite{vandermerwe_exoticfmcw_2022}. However, most PPDs seem to be generating CW and chirps \cite{vandermerwe_2023_ppd_comparison}, so we focus our analyses on these two classes, particularly on fast chirps that current tracking algorithms have a hard time catching. 

\subsection{Time-Domain Excision via a Complex First-Order IIR Notch}

The principal performance objective of any excisor is to maximize interferer suppression while minimally degrading GNSS \cite{borio_postcorr_degradation_2021}. This implies that, based on Eq. (\ref{eq:signal_model}), a good excisor needs to realize the inverse of $i[n]$ as closely as possible to cancel it \cite{eisfeller_2013_trackingsuppression_likeanc}. For narrowband interferers, one method of doing this with small latency and modest resource utilization is the complex one-pole IIR notch filter \cite{borio_2006_1polefilter}, defined as:
\begin{equation}
    H(z;k_a,z_0) \;=\; \frac{1 - z_0\,z^{-1}}{1 - k_a\,z_0\,z^{-1}},
    \label{eq:complex_notch}
\end{equation}
where $k_a\!\in\!(0,1)$ is the pole contraction factor determining the width of the notch, and the complex number $z_0[n]=\rho_z\,e^{j\theta_z[n]}$ determines the location and depth of the notch. The location, $f_z$, is determined by the angle of $z_0$, i.e., $f_z = \theta_z \cdot F_s / \, 2\pi$, with $\theta_z \in (-\pi,+\pi]$, and the magnitude, $\rho_z \in \Re^+$, which is typically unit magnitude or smaller, i.e., $\rho_z\!\le\!1$, controls notch depth \footnote{Note how Eq. \ref{eq:complex_notch} becomes an all-pass filter for $\rho_z = 0$, and $k_a\,\rho_z > 1$ causes unstable behavior \cite{ferdjallah_1994_anf} due to the IIR structure.}. However in most designs, $\rho_z = 1$, and a separate mechanism decides whether the notch filter is applied or not, simplifying the optimization landscape of the ANF to 1D. In other words, the problem becomes just finding $f_z$ for a unity-magnitude $z_0$, which runs faster than the 2D optimization for a full-complex $z_0$ that controls both depth and location \cite{gamba_2019_anfparametrization}. 

This simple notch filter is not the universally optimal excision mechanism. However, a comprehensive analysis of excisors, including the negative effects of notch filtering on code and carrier tracking loops of the downstream GNSS receiver \cite{gamba_postcorr_impact_2020}, is outside the scope of this paper, and we instead focus on the \emph{tracking mechanism} that localizes $i[n]$ using Eq. (\ref{eq:complex_notch}) as a signal model. Thus, the output $f_z[n]$ of the tracker we propose can also drive alternative excisors when they are preferable for better preserving GNSS structure (e.g., wavelet-domain attenuation reported to reduce GNSS distortion compared to a notch under similar conditions \cite{musumeci_wavelet_anf_2016}). Therefore, for the sake of simplicity and without loss of generality, we use the notch filter in Eq. (\ref{eq:complex_notch}) as the excisor in the rest of this paper, and focus on the frequency tracking problem.

\subsection{Instantaneous Frequency Tracking: Problem Statement and Error Metrics}

The tracking problem is characterized by the error metric $\varepsilon_f$ or equivalently $\varepsilon_\theta$:

\begin{equation}
   \varepsilon_f \;\triangleq\; \hat{f_z} - f_i ,
   \qquad
   \theta_i \triangleq 2\pi \frac{f_i}{F_s},
   \qquad 
   \varepsilon_\theta \;\triangleq\; \hat{\theta}_z - \theta_i \;=\; 2\pi\,\frac{\varepsilon_f}{F_s},
\end{equation}

where $\hat{f_z}$ denotes the frequency estimate produced by a tracker in Hz, and $\hat{\theta}_z$ is that in radians, respectively. Since we're evaluating frequency tracking in terms of how helpful it is for interference mitigation, a suppression metric should be derived from $\varepsilon_f$. It is possible to derive this for a static CW tone by evaluating Eq. \eqref{eq:complex_notch} at the interferer frequency $f_i$ and at a certain unit-circle (i.e., $\rho=1$) estimate of $z_0$, i.e., $\hat{z_0} = e^{j\hat{\theta}_z}$. However, this analytical expression would not be usable as is for chirp interferers because it implicitly assumes a linear time-invariant (LTI) notch, and the notch has to move for suppressing a chirp. Substituting the instantaneous $\varepsilon_\theta[n]$ for chirp interferers, especially fast sweeping ones, would wrongly estimate suppression amount since it would disregard the dynamics of the IIR filter for varying $\varepsilon_\theta[n]$. 

\begin{figure}[t]
  \centering
  \includegraphics[width=\textwidth]{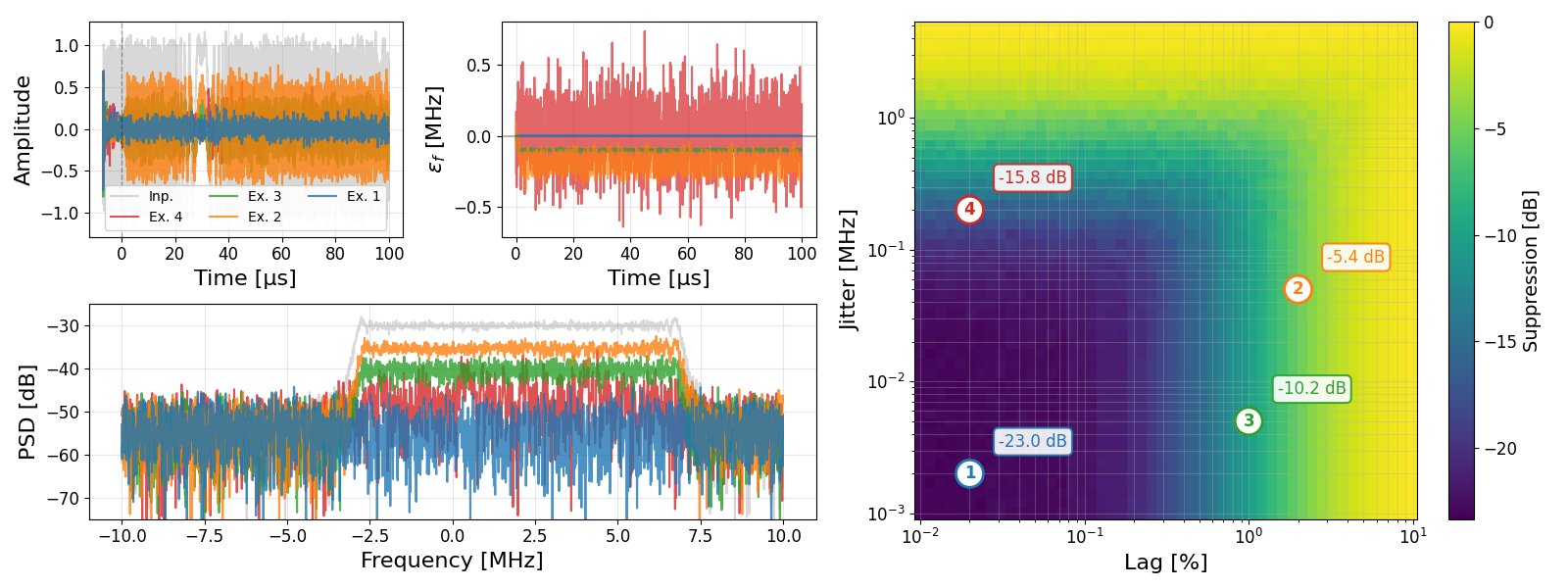}
  \caption{
  Simulated analysis of the effect of lag and jitter error on suppression performance for a chirp interferer sweeping 10~MHz in 100~\textmu s under a 20~dB jammer-to-noise power ratio. 
  (Top-left) Time-domain samples for the input (gray) and four representative example runs: Ex.~1 (blue, low lag--low jitter, near-ideal), Ex.~2 (orange, high lag--moderate jitter), Ex.~3 (green, high lag--lower jitter than Ex.~2), and Ex.~4 (pink, low lag--high jitter). 
  (Top-mid) Instantaneous frequency error~$\varepsilon_f[n]$ for the same examples. 
  (Bottom-left) Power spectral densities (PSDs) of the received signal before and after mitigation, color-coded as above. 
  (Right) Suppression heatmap (in~dB) versus lag (as~\% of chirp period, logarithmic) and steady-state jitter (MHz, logarithmic), with markers 1--4 corresponding to the example runs in the other plots.
  }  
  \label{fig:lag-jitter-panels}
\end{figure}

A generally valid suppression metric, applicable also to chirps, is the \emph{band-PSD ratio}: the ratio of the power spectral density (PSD) of the jammed input to that of the filtered output, computed over an evaluation window covering the jamming bandwidth. To show how this metric relates to $\varepsilon_f$, we run a small experiment in which we inject the two fundamental tracking error sources—\emph{lag} (i.e., low effective loop gain) and \emph{random jitter} (i.e., high gain)—onto the known instantaneous frequency of a chirp interferer and measure the resulting band-PSD ratio. Fig.~\ref{fig:lag-jitter-panels} maps this theoretical suppression versus these two errors for a representative case ($k_a{=}0.9$, $F_s{=}20$~MHz, a sweep from $-3$ MHz to $+7$~MHz over $100~\mu$s, JNR $20$~dB). The design objective of high suppression resides in the lower-left region of the plot (small lag, small jitter), whereas either excessive lag (right side) or excessive variance (top side) degrades suppression. Consequently, frequency trackers can be judged by how far their operating curve pushes toward this lower-left region. Our proposed Adam-ANF is designed exactly towards this target: adaptive-moment normalization raises effective gain during fast transients (reducing lag) while shrinking it at steady state (reducing jitter), thereby improving the speed–variance trade-off \cite{borio_loop_analysis_2016} that governs suppression performance.

%%%%%%%%%%%%%%%%%%%%%%%%%%%%%%%%%%%%%%%%%%%%%%%%%%%%%%%%%%%%%%%%%%%%%%%%%%%%%%%%%%%%%
%% Analytical derivation of suppression from notch eq for static CW only
%% we dropped this since it's not applicable to chirps
%% still keeping it commented here for reference
%% maybe an appendix?
% \begin{equation}
%     \bigl|H(z:=e^{j\theta_i};k_a,\hat{z_0})\bigr| 
% %    \;=\; \frac{\bigl|1 - e^{j\hat{\theta}_z} e^{-j\theta_i}\bigr|}{\bigl|1 - k_a e^{j\hat{\theta}_z} e^{-j\theta_i}\bigr|}
%     \;=\; \frac{\bigl|1 - e^{j\varepsilon_\theta}\bigr|}{\bigl|1 - k_a e^{j\varepsilon_\theta}\bigr|}
%     \;=\;
%     \frac{\sqrt{1+\rho^2 - 2\rho\cos\varepsilon_\theta}}
%          {\sqrt{1+(k_a\rho)^2 - 2k_a\rho\cos\varepsilon_\theta}}
% %    \;=\;
% %    \frac{\sqrt{2 - 2\cos\varepsilon_\theta}}
% %         {\sqrt{1+k_a^2 - 2k_a\cos\varepsilon_\theta}}
%     \;=\;
%     \frac{2\left|\sin\!\left(\tfrac{\varepsilon_\theta}{2}\right)\right|}
%          {\sqrt{(1-k_a)^2 + 4k_a\sin^2\!\left(\tfrac{\varepsilon_\theta}{2}\right)}},
%     \label{eq:atten_general}
% \end{equation}
%which should be minimized (i.e., $\varepsilon_\theta=0$) for maximum suppression. 
%%%%%%%%%%%%%%%%%%%%%%%%%%%%%%%%%%%%%%%%%%%%%%%%%%%%%%%%%%%%%%%%%%%%%%%%%%%%%%%%%%%%%

\section{Frequency Tracking with Adaptive Moments}
\label{sec:adam-anf}

Gradient-based frequency tracking methods share a common formalism: Once the filter structure is defined as in Eqs. (1-4), 
\begin{itemize}
    \item an objective function is chosen such that its minima occurs for the correct instantaneous frequency of the interferer,
    \item the gradient of this objective function with respect to the parameter controlling the notch frequency ($f_z$) is computed,
    \item an update rule that uses the gradient to recursively compute the value of that optimized parameter, is defined.
\end{itemize}
The canonical example of gradient-based trackers is the normalized-LMS (NLMS) method which uses the standard output power objective (MA+AR) of the one-pole complex IIR notch filter of Eq. \eqref{eq:complex_notch}, computes the gradient with respect to the $\theta_z$ parameter, and updates the value of the parameter with a nominal step size that is normalized by the input power arriving at the filter. The speed-variance trade-off of the method is thus spanned by the choice of the nominal step size. While various objective function selections are possible, these are orthogonal to our case since we are proposing a novel update rule, which can be used with any objective function. 

We first re-derive the NLMS solution in this formalism and then introduce our adaptive-moments update (Adam-ANF), which replaces the standard update rule with one that uses exponentially filtered first and second moments of the gradient.

\subsection{Objective Function and Gradient}
Let $x[n]$ be the input to the notch, and let $x_r[n]$ denote the AR output, i.e., latent variable of the filter, with the recursion
\begin{equation}
x_r[n] \;=\; x[n] + k_a\,z_0[n{-}1]\,x_r[n{-}1],
\qquad
z_0 \triangleq e^{j\theta_z}, \;\; k_a\!\in\!(0,1).
\end{equation}
The notch output then is
\begin{equation}
x_f[n] \;=\; x_r[n] - z_0[n{-}1]\,x_r[n{-}1].
\end{equation}
We minimize the instantaneous output (i.e., MA+AR) power
\begin{equation}
J[n] \;\triangleq\; |x_f[n]|^2,
\label{eq:obj}
\end{equation}
which is a useful proxy under the assumptions of additive white gaussian noise (AWGN) for $\eta[n]$ and for a useful GNSS signal that is much weaker than the interferer and the noise (i.e., $s_{GNSS}[n]$ is much weaker than $i[n]$ and $\eta[n]$).

To obtain an update in frequency, $f_z$ (1D), rather than directly in the $(\rho_z,f_z)$ 2D complex plane, we parameterize by the notch angle $\theta_z$ (and therefore the frequency $f_z = \theta_z F_s / 2\pi$) with $z_0[n]=e^{j\theta_z[n]}$. Using the chain rule,
\begin{equation}
g[n] \;\triangleq\; \frac{\partial J[n]}{\partial \theta_z}
\;=\;
2\,\Re\!\left\{\frac{\partial x_f[n]}{\partial \theta_z}\,x_f^*[n]\right\}
\;=\;
2\,\Re\!\Big\{-j\,z_0[n{-}1]\,x_r[n-1]\,x_f^*[n]\Big\},
\label{eq:grad-theta}
\end{equation}
where $^*$ denotes complex conjugate. Eq. \eqref{eq:grad-theta} is an unbiased stochastic gradient of the power objective Eq. \eqref{eq:obj} with respect to $\theta_z$, which is common to both the classical ANF method of NLMS as well as our proposed Adam-ANF since the two methods differ only by the update rule. 

\subsection{Classical ANF: Normalized LMS (NLMS)}
With Eq. \eqref{eq:grad-theta}, the classical normalized LMS update on the \emph{angle} reads
\begin{align}
\theta_z[n{+}1] \;&=\; \theta_z[n] \;-\; \mu[n]\;g[n],
\label{eq:nlms-theta}\\
\mu[n] \;&=\; \frac{\delta}{\widehat{E}\{|x_r[n]|^2\} + \epsilon_\mu},
\qquad 0<\delta\ll 1,
\end{align}
where $\widehat{E}\{\cdot\}$ is a causal power estimate (e.g., single-pole IIR, $\widehat{E}$ denotes sample-based expected value calculation) and $\epsilon_\mu$ is a small real number preventing division by zero. Intuitively, Eq. \eqref{eq:nlms-theta} increases loop gain when the AR state is small (i.e., low input power) and damps updates when the AR state is large (i.e., high input power), thereby normalizing the characteristics of the method across different JNR levels. The parameter $\delta$ controls the speed–variance trade-off: larger $\delta$ allows for faster tracking of chirps or sweeps but results in higher steady-state jitter, while smaller $\delta$ reduces jitter but slows down response to rapid frequency changes.

\subsection{Proposed Method: Adam-ANF}

We propose to replace the scalar step-size adaptation in Eq. \eqref{eq:nlms-theta} with \emph{adaptive moments} of the stochastic gradient, like in the popular neural network optimizer: Adam \cite{kingma2014adam}. We first justify this choice by describing the connections between ANFs and neural network literature, and then rigorously define the proposed Adam-ANF.

The intuition for this choice stems from the following: The notch filter of Eq. \eqref{eq:complex_notch} can be interpreted as a very simple shallow neural network with a single parameter to optimize, $f_z$. The network is in \emph{training mode} during normal operation of the filter, i.e., the input samples processed by the filter are training inputs, and the desired realization at the filter output is basically a value of 0 since the objective function is trying to minimize the output power as in Eq. \eqref{eq:obj}. In this formulation, the update rule in Eq. \eqref{eq:nlms-theta} is basically the \emph{optimizer}, which has been studied extensively in neural network literature as the update rule for its high-dimensional non-convex optimization problem of training \cite{schmidt_2021_optimizers}. Although exotic formulations have been shown to work well for different network structures depending on their depth, complexity and the types of operators they utilize or data they see during training, the Adam optimizer has been the most popular choice with overwhelming empirical evidence thus far. The reasons for its popularity are the simple formulation, fast convergence and low jitter around minima, which are exactly the desired properties for an ANF. 

Adam keeps exponentially weighted moving averages of the first and second moments, $m[n]$ and $v[n]$:
\begin{align}
m[n] &= \beta_1\,m[n{-}1] + (1{-}\beta_1)\,g[n], \\
v[n] &= \beta_2\,v[n{-}1] + (1{-}\beta_2)\,g^2[n],
\label{eq:adam}
\end{align}
with $m[{-}1]\!=\!0$, $v[{-}1]\!=\!0$, and two configuration parameters $0<\beta_1<1$ and $0<\beta_2<1$, where typically both values are chosen close to 0.99. Due to the initialization of the latent variables $m[n]$ and $v[n]$ from 0, the estimates are biased towards 0 at the start of the run, and the default Adam optimizer applies a correction term to alleviate this bias. This is not relevant for the ANF, which can run an indefinite number of steps, removing the bias itself naturally after a short time. Therefore, we do not apply the bias correction to reduce complexity, and run the recursion with:

\begin{equation}
\theta_z[n{+}1]
\;=\;
\theta_z[n]
\;-\;
\alpha\;\frac{m[n]}{\sqrt{v[n]} + \varepsilon_\alpha},
\label{eq:adam-update}
\end{equation}

where $\alpha$ is a base learning rate and $\varepsilon_\alpha$ is a small number that prevents division by zero. Note from Eqs. \eqref{eq:adam} and \eqref{eq:adam-update} that the AR power normalization feature in NLMS, or the similar self-tuning loop bandwidth terms employed in the adaptive FLL-based schemes \cite{vandermerwe_afllanf_2021}, is implicitly captured in Adam by $v[n]$, which normalizes the step with the inverse of gradient energy. In stationary conditions, $v[n]$ converges to the gradient variance and shrinks the effective step, reducing steady-state jitter; during fast chirps $v[n]$ remains low so the update acts with a larger effective gain, improving tracking speed. 

Moreover, the parameter $\beta_1$ controls the ``memory'' of the velocity estimate $m[n]$, or typically referred to as the momentum. Due to the purely stochastic nature of the gradient in the ANF context (updates are over each streamed sample, not in minibatches), the momentum term is not well-defined, so it only damps the response. For this reason we choose $\beta_1 \ll \beta_2$ to speed up convergence (typically $\beta_1=0.01~,~\beta_2=0.90$). In this configuration, which provides the best results for the ANF use case, the optimizer also resembles the lesser-known RMSProp method \cite{ruder2016optimizers}, which is basically a sub-class of the Adam optimizer that does not utilize the first moment at all. Similar to the NLMS, the speed-variance trade-off for the Adam-ANF is controlled by the base rate parameter $\alpha$, and $\beta_1$ and $\beta_2$ are typically fixed. 

\section{Performance Benchmark}
\label{sec:perf-benchmark}

In this section we benchmark the performance of the trackers in five stages: (i) simulations comparing NLMS-ANF and Adam-ANF across a range of JNR levels and chirp speeds, (ii) hardware verification over data recorded with the \textit{EDGE Microwave} test harness, (iii) runtime results on challenging recorded scenarios, (iv) a comparison with FLL-based methods, and (v) analyzing resource utilization as well as the effects of implementation constraints such as quantization and pipelining delays.

\subsection{Simulation-Based Comparison: NLMS-ANF vs.\ Adam-ANF}
\begin{figure}[h]
  \centering
  \includegraphics[width=\textwidth]{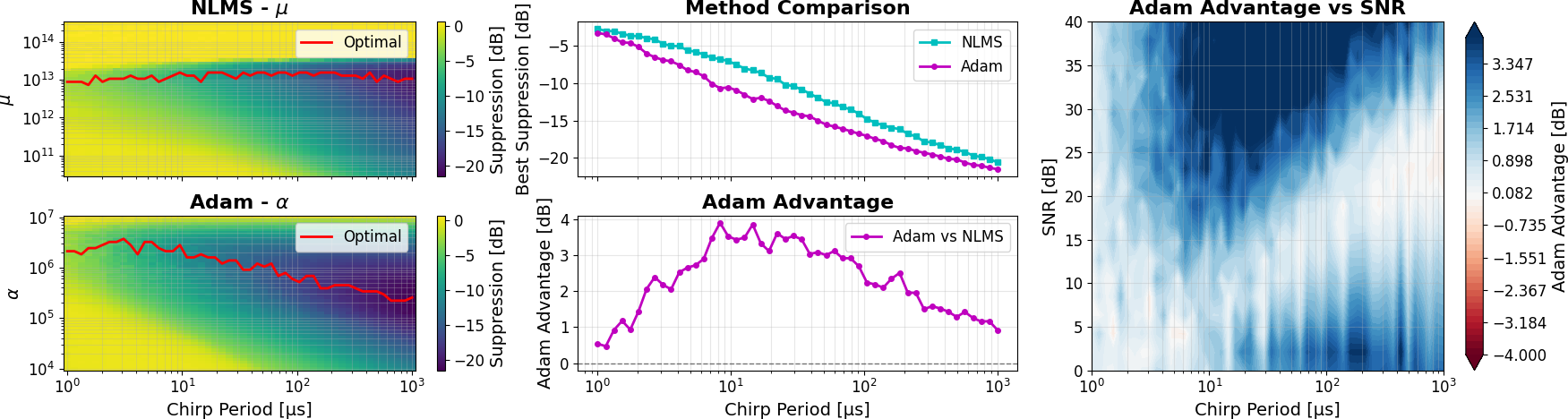}
  \caption{Gain parameter vs.\ chirp speed and chirp speed vs.\ JNR heatmaps show the suppression amount for chosen configurations. The red highlighted ridges mark Pareto-optimal operating points (i.e., optimal gain parameter per chirp speed in terms of best suppression).}
  \label{fig:heatmap-speed-param}
\end{figure}

We simulate the performance of the two methods over a wide range of chirp speeds and jammer-to-noise ratio (JNR) levels. The left heatmap and line plots in Fig. \ref{fig:heatmap-speed-param} show the best attainable performance from each method for different chirp speeds: We sweep the principal gain hyperparameters of each method and, for each chirp speed, identify the Pareto-optimal operating point (maximum suppression), averaged over [0, 40] dB JNR. Adam-ANF provides an advantage of up to 4 dB over NLMS-ANF, varying with respect to the sweep speed, as well as a wider safe tuning range (blue region) for the gain parameter.

To quantify the performance under various power levels, the right contour plot in Fig. \ref{fig:heatmap-speed-param} also sweeps against [0, 40] dB JNR. Results show that the advantage persists in high JNR levels and faster chirp speeds, but the extremely fast chirps and/or low JNR resets the advantage, causing the two methods to perform similarly. In summary, the maximum advantage of Adam-ANF over NLMS-ANF is shown to be approximately 4 dB, decreasing for the fastest chirps or lowest JNR levels.

\subsection{Experimental Verification: NLMS-ANF vs.\ Adam-ANF on Recordings}
\begin{figure}[h]
  \centering
  \includegraphics[width=\textwidth]{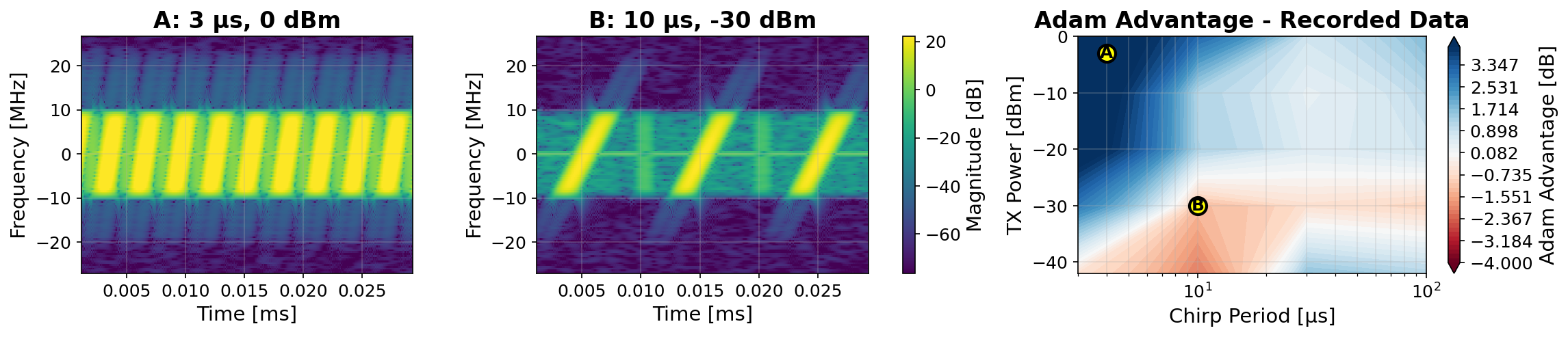}
  \caption{Experimental verification of Fig. \ref{fig:heatmap-speed-param} with recorded data at various JNR and chirp speed levels. Input spectograms at various points on the contour are also shown (note the effect of the 20 MHz lowpass filter on the RF frontend).}
  \label{fig:heatmap-speed-param-exp}
\end{figure}

We repeat the same performance comparison on data recorded with an in-house \textit{EDGE Microwave} test harness\footnote{All necessary procedural precautions were taken so the PPD emission could not couple into or degrade any receiver outside the test enclosure.}, evaluating the performance advantage of the Adam-ANF over the NLMS-ANF for a 50 MHz-wide chirp, at chirp periods of $\{3,10,30,100\}\ \mu s$ and transmit power levels of $\{-42,-30,-20,-10,0\}\ \text{dBm}$. The results in Fig. \ref{fig:heatmap-speed-param-exp} verify the simulated comparison, showing approximately a 4 dB maximum advantage of Adam-ANF over NLMS-ANF. Moreover, the disappearance of the Adam-ANF advantage for the fastest chirps, and/or extremely low JNR, is also demonstrated in these experiments, as seen by the contour plot in Fig. \ref{fig:heatmap-speed-param-exp} having similar shape to that in Fig. \ref{fig:heatmap-speed-param}.

\subsection{Runtime Results on Challenging Scenarios}
We next compare Adam-ANF to a classical NLMS-ANF on two challenging recorded RFI events: a capture of a commercially available fast PPD on the \textit{EDGE Microwave} test harness, and a capture of a generated low JNR linear wideband chirp from the GNSS Interference dataset of the ATLANTIS Horizon EU Project by LINKS Foundation \cite{mehr_2025_atlantis_dataset}. For each recording, we show spectrograms, time–domain responses, and pre/post-excision PSDs over the interferer band, where suppression is reported as the band-PSD ratio as defined in Section~II.3. Both methods drive the same one-pole complex IIR in notch Eq.~\eqref{eq:complex_notch}, updates occur at every clock cycle, and we tune the gain parameter of each tracker per recording to record its best performance for that setting, for fair comparison. 

\begin{figure}[h]
  \centering
  \includegraphics[width=\textwidth]{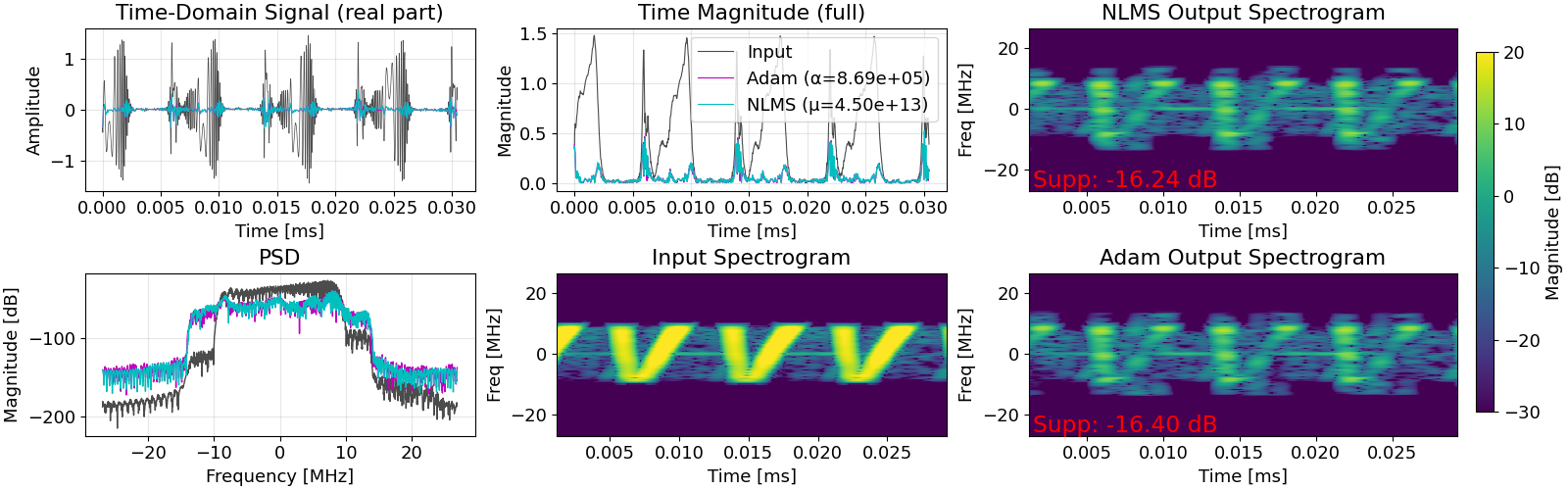}
  \caption{Recorded COTS PPD: time-amplitude and time-magnitude plots, spectrogram, and PSD before/after excision. The difference between Adam-ANF and NLMS is smaller, $\approx$ 0.2 dB, but still improving suppression. The spectogram color representing the minimum power is tuned to the in-band ($\pm10$ MHz) noise floor. Both methods fail to suppress the jammer down to the noise floor.}
  \label{fig:edge-recorded-tip2}
\end{figure}

Fig. \ref{fig:edge-recorded-tip2} show the response of both Adam-ANF and NLMS-ANF to RFI from a PPDs. The test harness is tuned to center the wideband jammer, making fast ramp tracking the main performance attribute. Results show that both methods struggle tracking the very fast jammer, and the advantage of the Adam-ANF over NLMS-ANF is minimal, around 0.2 dB, as expected by the analyses of the previous subsection. 

\begin{figure}[h]
  \centering
  \includegraphics[width=\textwidth]{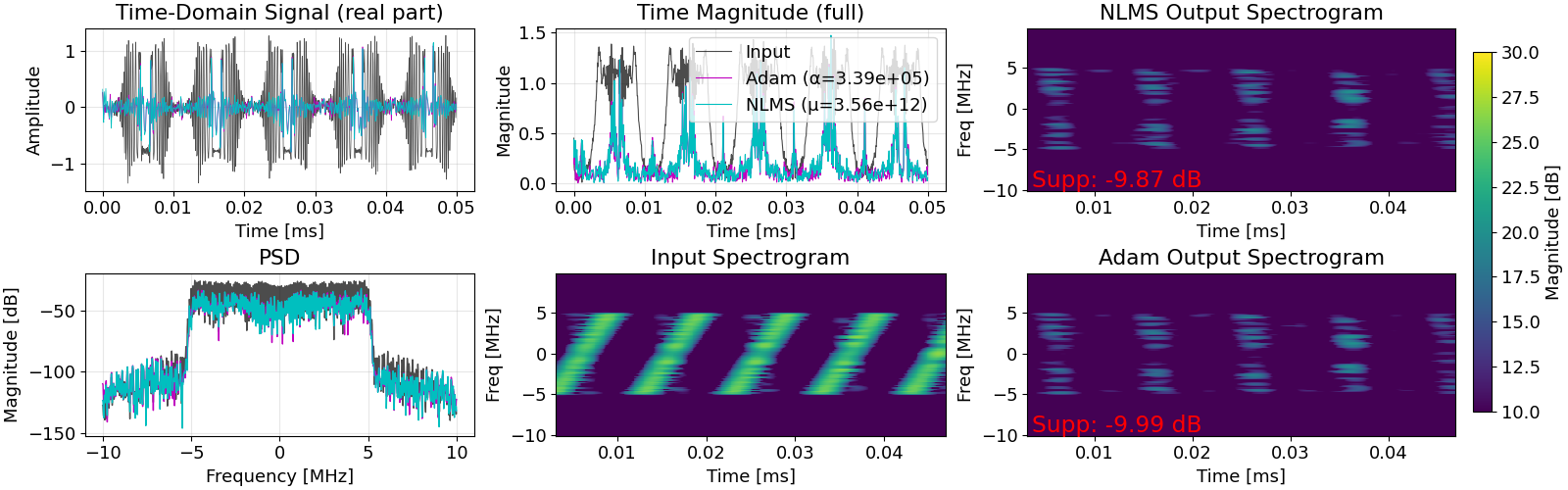}
  \caption{Recorded chirp from the ATLANTIS dataset: time–amplitude and time-magnitude plots, spectogram, and PSD before/after excision. Adam-ANF is only marginally better since the chirp is low JNR. The spectogram color representing the minimum power is tuned to the in-band ($\pm5$ MHz) noise floor. Both methods suppress the jammer down to the noise floor except for the chirp reset discontinuities.}
  \label{fig:atlantis-recorded}
\end{figure}

We next repeat the experiment for a recording of a linear wide chirp from the publicly available ATLANTIS dataset \cite{mehr_2025_atlantis_dataset}. Fig. \ref{fig:atlantis-recorded} shows the performance of both methods against this linear wide chirp. Although the chirp is not extremely fast (10 MHz in less than 10 $\mu$s), the two methods again perform similarly here since the scenario is a low JNR one, but both methods successfully suppress the jammer. This supports the analysis in the previous subsection, where Adam-ANF advantage over NLMS-ANF diminished against extremely fast chirps and low JNR. 

\subsection{Comparison with FLL Baselines}
\label{subsec:fll-vs-others-sim}

\begin{figure}[h]
  \centering
  \includegraphics[width=\textwidth]{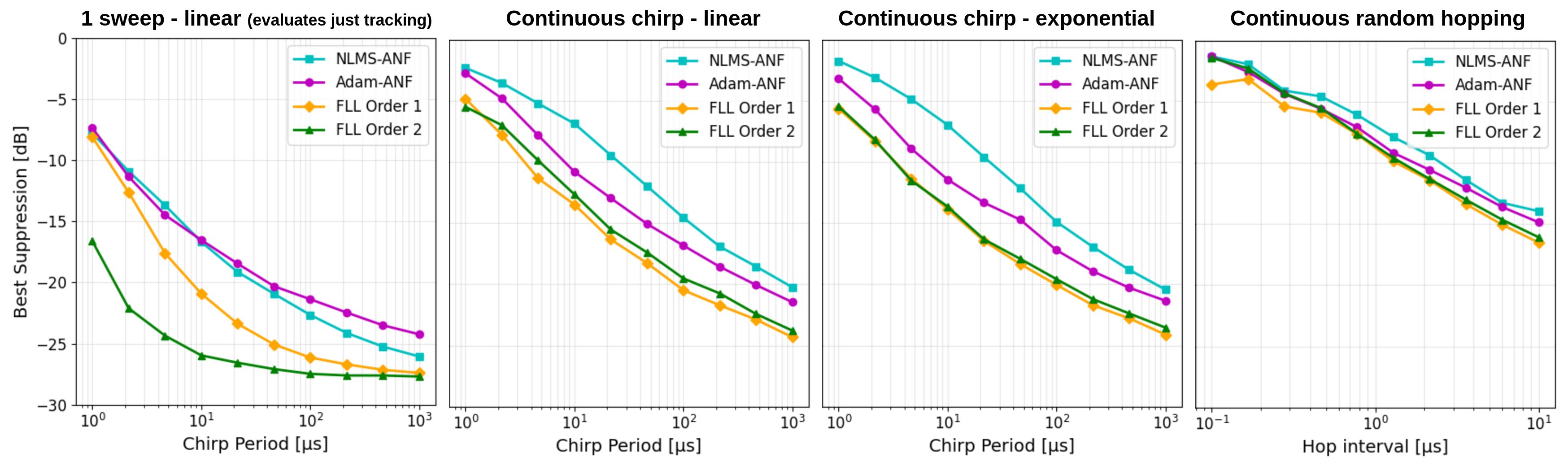}
  \caption{Best-achieved suppression (dB) versus chirp period (or hop interval) for NLMS-ANF, Adam-ANF, and type-I/type-II FLLs under single-sweep, continuous linear chirp, continuous exponential chirp, and continuous random hopping.}
  \label{fig:suppression-speed}
\end{figure}

We compare the performance of first- and second-order FLLs described in \cite{gamba_2018_fll4chirp_anf_analysis} with NLMS-ANF and Adam-ANF across four simulated scenarios: a single linear sweep, a continuous linear chirp, a continuous exponential chirp, and a continuous random hopping process. For each chirp period, the per-method gain hyperparameter is optimized for the best-achieved band–PSD suppression, following the same procedure from earlier runs. The results are summarized in Fig.~\ref{fig:suppression-speed}.

The single linear sweep purely characterizes tracking performance since there are no discontinuities. The second order FLL attains the highest suppression across the range of sweep periods, consistent with the zero steady-state slope error property of type-II loops for linear chirps \cite{gamba_2018_fll4chirp_anf_analysis}, i.e., the second order FLLs are directly modeled for linear chirps. The first order FLL performs worse than the second order FLL and, for short periods, approaches the performance of the gradient-based methods (short periods, i.e., faster chirps, are closer to discontinuities). Adam-ANF yields a small improvement over NLMS-ANF, but since FLL-based methods are basically model-based, it's inferior to both FLLs.

For the continuous linear chirp, both FLL variants outperform the ANFs over most chirp speeds. The gap between first and second order FLLs narrows relative to the single-sweep case since the discontinuities dominate the suppression performance (note how overall performance is also lower). Adam-ANF now dominates NLMS-ANF since the adaptive properties of Adam-ANF catches discontinuous jumps faster. For the continuous exponential chirp, which is piece-wise linear for slower chirps, the relative ordering is similar, with the FLLs maintaining the advantage. Adam-ANF and NLMS-ANF are still below the FLLs across both chirps, but Adam-ANF's advantage over NLMS-ANF is still present. For the continuous random hopping case, all methods exhibit reduced suppression compared to chirps since the signal is now only made of discontinuities. The lack of a predictable slope eliminates the structural advantage of second order loops, and gradient methods cannot anticipate abrupt jumps. As a result, all methods converge to broadly similar, overall degraded performance levels in this regime.

Overall, these experiments indicate that Adam-ANF offers significant improvement over NLMS-ANF, proving that it's a better gradient-based method. However, it does not outperform FLLs, which provide higher suppression than gradient-based methods when the interferer dynamics are aligned with the loop model (tones and chirps). This aligns well with the theoretical expectation since FLLs are modeled directly with respect to such type-I and type-II signals where gradient based methods make no assumptions on the data, thus facing harder convergence dynamics. We discuss the utility of this \emph{model-free} property of gradient-based methods in the conclusion for future work.

\subsection{Resource Utilization, Quantization and Pipelining Delays}
\label{sec:impl}

\noindent We finally evaluate resource utilization as well as performance hits under realistic digital constraints, i.e., finite precision and pipelining latency. We run a representative scenario (10~MHz sweep over $100~\mu$s, $F_s\!=\!20$~MHz, $k_a\!=\!0.70$) and average the results over [0, 40] dB JNR. Each tracker (NLMS-ANF, Adam-ANF, FLL~order~1\&2) is first tuned to its optimal gain hyperparameter for this scenario, and then the implementation constraints are introduced at varying levels. Specifically, fixed-point quantization of the algorithm at increasing bit levels between 4 and 20 bits, as well as $z^{-D}$ pipelining delays on the estimate-to-filter loop, are applied. Suppression is reported as the ratio of in-band PSDs (pre/post excision) over the chirp band.

\begin{figure}[t]
    \centering
    \includegraphics[width=\textwidth]{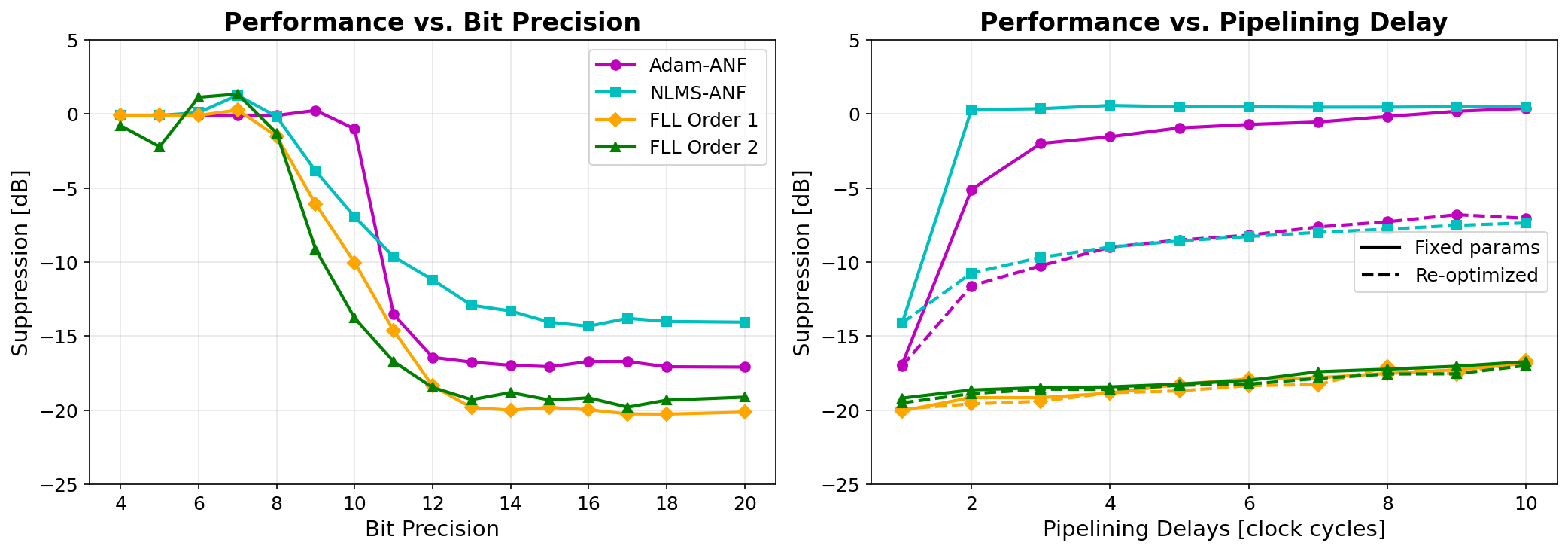}
    \caption{Implementation analysis. Left: suppression vs.\ bit precision when the estimator and notch operate in fixed-point. Right: suppression vs.\ estimator control-path latency (clock-cycle delay). Each curve is the JNR-averaged result with method hyper-parameters either fixed to their full-precision zero-delay optima, or re-optimized for each delay value.}
    \label{fig:impl}
    \end{figure}

\noindent Fig. \ref{fig:impl} shows that all methods exhibit a sharp degradation below $\approx\!8$~bits, and at $\geq\!14$~bits performance approaches the floating-point baseline. Adam-ANF degrades more gracefully than NLMS-ANF starting from 10 bits. The type-II FLL attains the highest absolute suppression at high precision but shows a clearer “knee” too as precision is reduced. 

Moreover, added pipelining delay monotonically reduces suppression. Gradient trackers cannot tolerate added pipelining delay with the same gain hyperparameter. The FLLs are more robust against the pipelining delay compared to the gradient-based methods, however, their performance also suffers up to a 4 dB loss. Another experimental observation is that the optimal gain of the gradient methods tightly depend on pipeline delays. Therefore if deeper pipelines are unavoidable on hardware, retuning loop gains (Adam-ANF's $\alpha$ and NLMS-ANF's $\mu$) with quantization and delay effects recovers part of the loss (see dashed lines on the figure). However, as a verification of the theoretical understanding in this area, it's evident that the gradient-based methods suffer significantly higher losses against pipelining delay in comparison to the FLL-based approaches.

\noindent The resource utilization summary of all four approaches are summarized in Table \ref{tab:per_sample_ops}. 

\begin{table}[h]
\centering
\setlength{\tabcolsep}{8pt}
\renewcommand{\arraystretch}{1.1}
\caption{Representative per-sample real-operation counts for the four trackers when driving the same one-pole complex IIR notch of Eq.~(\ref{eq:complex_notch}). Counts assume reuse of $z_0[n{-}1]\,x_r[n{-}1]$ between $x_r$ and $x_f$, and standard real-operation decompositions for complex arithmetic (one complex multiply $\rightarrow$ 4 mult + 2 add). ``trig'' denotes the \emph{per-sample} evaluation of $\sin$/$\cos$ (or equivalent CORDIC) for $z_0[n]\!=\!e^{j\theta_z[n]}$.}
\label{tab:per_sample_ops}
\begin{tabular}{lcccccc}
\hline
\textbf{Method} & \textbf{Add} & \textbf{Sub} & \textbf{Mult} & \textbf{Div} & \textbf{$\sqrt{\cdot}$} & \textbf{Trig}\\
\hline
NLMS-ANF & 9 & 3 & 16 & 1 & 0 & 1\\
Adam-ANF (proposed) & 8 & 3 & 17 & 1 & 1 & 1\\
FLL (1st order) & 8 & 2 & 11 & 0 & 0 & 2\\
FLL (2nd order) & 10 & 2 & 12 & 0 & 0 & 2\\
\hline
\end{tabular}
\end{table}

\noindent All methods share the same notch core: computing $x_r[n]=x[n]+k_a z_0[n{-}1]x_r[n{-}1]$ and $x_f[n]=x_r[n]-z_0[n{-}1]x_r[n{-}1]$ (one reused complex product), which contributes 6 multiplies, 4 additions, and 2 subtractions. Gradient-based methods (NLMS-ANF and Adam-ANF) add one extra complex product for $g[n]$ (Eq.~\eqref{eq:grad-theta}). NLMS additionally needs a power estimator and a division for the normalized step, whereas Adam-ANF needs one square-root and one division operation for the adaptive-moment normalization (Eqs.~\eqref{eq:adam}--\eqref{eq:adam-update}). FLLs use a phase-difference discriminator via $x[n]x^*[n{-}1]$ and a trigonometric operation, i.e., $\operatorname{atan2}$, followed by a loop filter. Second-order FLL adds one more state update (extra mult+adds). All methods compute $z_0[n]=e^{j\theta_z[n]}$ once per sample.

\vspace{10px}

\noindent In summary, the proposed Adam-ANF provides higher performance than NLMS-ANF at the cost of an additional multiplication operation and a square-root operation. However, the FLL-based methods prove to be more advantageous in terms of resource utilization too, despite an additional trigonometric operation. This highlights the utility of using signal-model based approaches, when such a signal model is available.  

\section{Conclusion}

\noindent This paper presents an adaptive-moment update (Adam-ANF) for instantaneous frequency tracking in narrowband GNSS interference mitigation. The proposed method preserves the standard one-pole complex IIR notch and output power minimization objective and replaces the classical NLMS step with exponentially weighted first and second moments of the stochastic gradient.

\noindent Across simulated and recorded RFI events, Adam-ANF is shown to improve the speed–variance trade-off of gradient-based tracking. Relative to NLMS-ANF, it yields up to 4~dB higher in-band suppression and a wider safe tuning range under moderate-to-high JNR and a broad span of chirp rates. This advantage dminishes for extremely fast sweeps and very low JNR. Comparisons with FLL baselines confirm that first- and second-order FLLs remain superior for linear sweeps and tones, which are the types of signal they are modeled for, while Adam-ANF consistently outperforms NLMS-ANF across all dynamics. Implementation analyses reveal sharp degradation below about 8~bits of quantization and sensitivity to estimator-to-filter pipelining latencies, but downscaling the loop gain partly recovers performance penalties. Type-II FLLs achieve the highest absolute suppression at high precision, and Adam-ANF is slightly more gracious than NLMS-ANF under quantization. 

\noindent While FLL-type mitigators are providing a better overall performance than gradient-based ANFs, the difference reduces in the face of discontinuities like chirp resets or random hops, which are the two cases where the modeling assumptions of FLLs break down. This highlights another promising future area of research on gradient-based methods, which are higher-dimensional parametrizations against more complex interferers and multiple simultaneous sources. The Adam-ANF naturally scales to such higher-dimensional parameterizations (e.g., multi-notch or structured all-pass cascades) because it uses the Adam optimizer which has a proven track record in training deep neural networks with billions of parameters. This suggests “adaptive neural filters” (ANeF) with multiple coupled parameters for (i) simultaneous tracking of multiple jammers, (ii) joint frequency–slope–depth adaptation, or (iii) hybrid time/frequency-domain excisors sharing a common objective like power minimization. In contrast, discriminator-based FLL loops scale poorly with dimensionality due to loop interaction and gain scheduling complexity. A systematic exploration of such ANeF designs, stability regions under finite precision and pipelining delays, and joint detectors for notch gating, is left open for future work.

%The Adam-ANF update rule naturally scales to higher-dimensional parameterizations (e.g., multi-notch or structured all-pass cascades) because it uses the Adam optimizer which has a proven track record in training deep neural networks with billions of parameters. This suggests “adaptive neural filters” (ANeF) with multiple coupled parameters for (i) simultaneous tracking of multiple jammers, (ii) joint frequency–slope–depth adaptation, or (iii) hybrid time/frequency-domain excisors sharing a common objective like power minimization. In contrast, discriminator-based FLL loops scale poorly with dimensionality due to loop interaction and gain scheduling complexity. A systematic exploration of ANeF designs, stability regions under finite precision, and joint detectors for notch gating is left open for future work.

\section*{acknowledgements}

\noindent The authors thank the \textit{EDGE Microwave} team for their contributions to the development and testing of the system. In particular, we are grateful to Yusuf \c{S}ahin and Semih \.{I}nci for their support with hardware integration, test harness operation, data collection, and many helpful discussions.

% the apacite bibliography style matches the ION bibliography style guidelines.
\bibliographystyle{apalike}
\bibliography{references}

\end{document}